\documentclass[twocolumn,preprintnumbers,amsmath,amssymb,10pt,a4paper,showkeys,superscriptaddress]{revtex4}
\usepackage{graphicx} \usepackage{bm}
\usepackage{amsmath,amsfonts,amssymb}

\usepackage{geometry}
\usepackage{color} 

\begin{document}
\title{General analysis of mathematical models for bone remodeling}
\author{Martin Zumsande}
\email{zumsande@mpipks-dresden.mpg.de}
\affiliation{Max Planck Institute for the Physics of Complex Systems, 01187 Dresden, Germany.} 
\author{Dirk Stiefs}
\affiliation{Max Planck Institute for the Physics of Complex Systems, 01187 Dresden, Germany.} 
\author{Stefan Siegmund}
\affiliation{Department of Mathematics, Dresden University of Technology, 01062 Dresden, Germany.}
\author{Thilo Gross}
\affiliation{Max Planck Institute for the Physics of Complex Systems, 01187 Dresden, Germany.} 
\date{\today}

\begin{abstract}
Bone remodeling is regulated by pathways controlling the interplay of osteoblasts and osteoclasts.
In this work, we apply the method of generalized modelling to systematically analyse a large class of models of bone remodeling.
Our analysis shows that osteoblast precursors can play an important role in the regulation of bone remodeling.
Further, we find that the parameter regime most likely realized in nature lies close to bifurcation lines, marking qualitative changes in the dynamics.
Although proximity to a bifurcation facilitates adaptive responses to changing external conditions, it entails the danger of losing dynamical stability. Some evidence implicates such dynamical transitions as a potential mechanism leading to forms of Paget's disease.
\end{abstract}


\maketitle

\section{Introduction}
Bone is a complex tissue that is being repaired and rebuilt continuously throughout an individual's life. The process of bone remodeling consists of two subprocesses: the resorption of old bone and the formation of new bone.
In the past decades it has become clear that, at the cellular level, bone remodeling depends on the interplay between two different types of cells, osteoclasts and osteoblasts. The former are multinuclear cells of hematopoietic origin that resorb bone, while the latter are mononuclear cells of mesenchymal origin that fill the gaps left by osteoclasts with newly formed bone tissue \cite{Manolagas00, Robling06}.

From a theoretical perspective, bone remodeling is interesting because biological requirements seem to pose contradictory demands. On the one hand, the system must show robustness with respect to naturally occurring fluctuations. On the other hand, the system must show adaptivity to relevant changes in the external conditions that require an increased or decreased rate of bone remodeling.

In humans, the process of bone remodeling is regulated by several autocrine and paracrine factors to maintain the balance of bone.
In particular, it has been discovered that a signaling pathway involving the Receptor Activator of NF-$\mathrm{\kappa}$B (RANK), its ligand RANKL, and the cytokine osteoprotegerin (OPG) play an important role in the regulation of bone remodeling \cite{Anandarajah09,Boyce08}. For osteoclasts to mature, it is necessary that RANKL, expressed by cells of osteoblastic lineage, attaches to RANK, expressed on cells of osteoclastic lineage. This process is regulated by the decoy receptor OPG, which is expressed by osteoblastic cells and inhibits the differentiation of osteoclasts by binding to RANKL and thus sequestering it.
Another important regulator, the cytokine TGF$\beta$, is known to influence both osteoclasts and osteoblasts \cite{Janssens05}. Over- or underexpression of TGF$\beta$ and the protagonists of the RANKL pathway is related to several diseases of bone, such as osteoporosis and Paget's disease of bone \cite{Reddy01,Whyte06,Kearns08,McNamara10}.

At a particular site, osteoblasts and osteoclasts move through bone in a group, remodeling tissue on its way. Such a collection of cells is known as a \emph{basic multicellular unit} (BMU) \cite{Frost69}. The dynamics of a single BMU is difficult to model because spatial aspects become crucial. In this study, mathematical models of ordinary differential equations (ODEs) are studied, which is an approximation that has often been applied in earlier studies. The justification for the omission of spatial effects is that an average is taken over many BMUs, thus analysing systemic properties of bone remodeling. 

Mathematical models of bone remodeling were studied in various earlier works. In particular, a minimal model consisting of two ODEs was constructed in Ref.~\cite{Komarova03}. In this model, the terms describing the regulation were assumed to be power laws. Thereby, the effects of paracrine and autocrine factors were condensed into power law exponents.
A more detailed model was formulated in \cite{Lemaire04,Pivonka08,Pivonka10} which incorporated three dynamic variables and made use of Michaelis-Menten kinetics.

The earlier works pointed out that under physiological conditions and in absence of external stimuli, the system should reside in a steady state, where the numbers of osteoclasts and osteoblasts remain approximately constant in time. 
For the system to remain close to the steady state, the state has to be dynamically stable, so that the system is driven back to the steady state after small perturbations. At the same time it is desirable that the stationary densities of osteoblasts and osteoclasts react sensitively to external influences, communicated through the signalling molecules. Mathematically speaking this means that the system should be robust against fluctuations of the variables, but sensitive to changes in the parameters. In dynamical systems, the strongest response of steady states to paramater change is often found close to bifurcations -- critical thresholds at which the stability to perturbations is lost.
Therefore, it is intuitive that there should be some tradeoff between dynamical stability and responsiveness.
It is thus possible that the physiological state of the bone remodeling system is characterized by parameter values close to a bifurcation point. 

An intriguing possibility, raised in \cite{Komarova03}, is that some diseases of bone might have their cause not in a shift of the steady-state concentrations, but in a bifurcation, in which the stability of the steady state is lost.
Dynamical systems theory has established a large variety of powerful tools for detecting and analyzing bifurcations. If a given disease were found to be related to a bifurcation phenomenon, this arsenal of tools could be utilized for understanding the causes and consequences of the disease.

In dynamical systems, the dynamics close to steady states are governed by the Jacobian matrix \cite{Guckenheimer97,Kuznetsov95}. 
In a $N$-dimensional (i.e., $N$-variable) system the Jacobian has $N$ eigenvalues which are either real, or form complex-conjugate eigenvalue pairs.
Local bifurcations occur when a change of parameters causes one or more eigenvalues of the Jacobian matrix to cross the imaginary axis, so that an eigenvalue with a negative real part becomes an eigenvalue with a positive real part. This can occur in two fundamental scenarios: In the case of a \emph{saddle-node bifurcation}, a real eigenvalue crosses the imaginary axis and becomes positive. This bifurcation typically occurs at a threshold at which steady states collide and vanish, which can lead to abrupt transitions in the system. 
In the case of the \emph{Hopf bifurcation}, a complex conjugate pair of eigenvalues crosses the imaginary axis, acquiring positive real parts, while the stability of the steady state is lost. We can distinguish between two types of Hopf bifurcations: In a \emph{supercritical Hopf bifurcation}, a stable limit cycle is born, leading to small-amplitude sustained oscillations. In a \emph{subcritical Hopf bifurcation}, an unstable limit cycle that coexists with a stable steady state vanishes, typically leading to a catastrophic loss of stability and, often, large-amplitude oscillations \cite{Kuznetsov95}.

In the present work, we explore a large class of mathematical models for the regulation of bone remodeling with respect to their bifurcation properties. To this end, we apply the method of generalized modeling \cite{Gross06, Steuer06, Gross09, Zumsande10} which allows analyzing models in which the reaction kinetics is not restricted to specific mathematical functions.
Thereby, generalized models can provide a broad overview of the dynamics of the system, which facilitates the choice of specific models. More importantly, the analysis of the generalized model reveals dynamical instabilities that are potentially related to pathologies of bone remodeling.

The models proposed in the present paper generalize and extend findings from earlier specific models.
Specifically, we find that among the previously discussed scenarios for regulatory interactions the one that is known to yield the highest responsiveness leads to steady states close to bifurcations. Crossing these bifurcations causes trajectories to leave the dynamical regime of healthy bone remodeling and can possibly be related to diseases of bone. We show that in two-variable models with the structure proposed in \cite{Komarova03}, stability of the steady state requires that the actions of OPG dominate over those of RANKL, while three-dimensional models \cite{Lemaire04,Pivonka08} can also be stable without that assumption. These results suggest that osteoblast precursors have an important impact on the dynamics and should be taken into account explicitly in models.

\section{Model construction}

A mathematical model capable of describing the process of bone remodeling must take into account the concentrations of active osteoblasts, $B$, and of active osteoclasts, $C$. We start by discussing a minimal model of these two dynamic variables in Sec.~\ref{sec:2varintro}.
Because the minimal model potentially oversimplifies the problem by ignoring the populations of precursor cells, especially in the case of osteoblasts, a more detailed model that accounts for osteoblast behavior at different stages of maturation is introduced in Sec.~\ref{sec:3varintro}.

\subsection{Two-variable model}
\label{sec:2varintro}
In the minimal two-variable model, we assign to both state variables a gain term ($F$,$H$, respectively) describing the maturation of new cells from a pool of precursors, and a loss term ($G$,$K$) describing the removal of cells due to death or further differentiation. This leads to the basic equation system
\begin{equation}
\begin{split}
\frac{\mathrm{d}}{\mathrm{d} \mathrm{t}} B &= F(B,C) - G(B,C) \\
\frac{\mathrm{d}}{\mathrm{d} \mathrm{t}} C &= H(B,C) - K(B,C)
\end{split}.
\label{eq:generalset}
\end{equation}
In the following, we assume that the functions $F(B,C)$,$G(B,C)$,$H(B,C)$ and $K(B,C)$ are positive and continuously differentiable, but do not restrict them to specific functional forms.

Our aim is to derive mathematical conditions on the stability of all possible steady states in all models of the form introduced in (\ref{eq:generalset}). As we show below, the threshold parameter values at which the stability changes (i.e., the bifurcation points) can be expressed as a function of parameters, which have a clear biological interpretation in the context of bone remodeling. 

In any particular steady state we can denote the number of osteoblasts and osteoclasts as $B^*$ and $C^*$, respectively. We can then define normalized variables 
\begin{equation}
b=\frac{B}{B^*}, \quad c=\frac{C}{C^*}.
\end{equation}
Similarly, we define a set of normalized functions
\begin{equation}
\begin{split}
&f(b,c)=\frac{F(B,C)}{F^*(B^*,C^*)}, \quad g(b,c)=\frac{G(B,C)}{G^*(B^*,C^*)}, \\
&h(b,c)=\frac{H(B,C)}{H^*(B^*,C^*)}, \quad k(b,c)=\frac{K(B,C)}{K^*(B^*,C^*)}.
\end{split}
\end{equation}
Using these definitions, the system can be written as
\begin{equation}
\begin{split}
&\frac{\mathrm{d}}{\mathrm{d} \mathrm{t}} b = \alpha_1 \left( f(b,c) - g(b,c) \right) \\
&\frac{\mathrm{d}}{\mathrm{d} \mathrm{t}} c = \alpha_2 \left( h(b,c) - k(b,c) \right).
\end{split}
\label{eq:generalsetnorm}
\end{equation}
where 
\begin{equation}
\alpha_1=\frac{F^*(B^*,C^*)}{B^*}=\frac{G^*(B^*,C^*)}{B^*}
\label{eq:alp1}
\end{equation}
and 
\begin{equation}
\alpha_2=\frac{H^*(B^*,C^*)}{C^*}=\frac{K^*(B^*,C^*)}{C^*}.
\label{eq:alp2}
\end{equation}
The second equalities in Eq.(\ref{eq:alp1}) and Eq.(\ref{eq:alp2}) hold because gain and loss terms have to balance in a steady state.

In the new coordinates, the formerly unknown steady state is located at $(b,c)=(1,1)$. 
As mentioned in the introduction, this steady state is said to be asymptotically stable if the system returns to it after a sufficiently small perturbation \cite{Kuznetsov95,Guckenheimer97}. This is the case if all eigenvalues of the Jacobian matrix have negative real parts.
In the present model the Jacobian can be written as
\begin{equation}
\label{eq:jacobian2d}
\rm \bf J=\begin{pmatrix} \alpha_1 & 0 \\ 0 & \alpha_2  \end{pmatrix}
\begin{pmatrix} 
f_{\rm b}-g_{\rm b} & f_{\rm c}-g_{\rm c} \\ 
h_{\rm b}-k_{\rm b} & h_{\rm c}-k_{\rm c} \end{pmatrix}.
\end{equation}
Here we used Roman subscripts to indicate partial derivatives. For instance, $f_{\rm b}$ is defined as
\begin{equation}
\begin{split}
&f_{\rm b}=\left. \frac{\partial\mspace{2mu} f}{\partial\mspace{2mu} b}\right|_{b=1,c=1}= \left. \frac{B^*}{F^*} \frac{\partial\mspace{2mu} F}{\partial\mspace{2mu} B}\right|_{B=B^*,C=C^*}\\
&= \left. \frac{\partial\mspace{2mu} \left( \ln F \right)}{\partial\mspace{2mu} \left( \ln B \right)}\right|_{B=B^*,C=C^*}.
\end{split}
\label{eq:elastidef}
\end{equation}

So far we succeeded in constructing the Jacobian matrices corresponding to steady states in a very general class of models. We emphasize that we did not have to assume that there was only one steady state in the system. In the general case where more steady states exist, the formal derivation of the Jacobian applies to all steady states in all models within the class considered here, whereas the quantities appearing in the Jacobian matrix generally differ between steady states. Although these quantities, such as $f_{\rm b}$, are in general unknown, they do not depend on the dynamical variables and can therefore be treated as unknown parameters with the same right as the parameters that are introduced in conventional models. 
Just as conventional parameters, the generalized parameters appearing in the Jacobian have a well-defined interpretation in the context of the model. In the remainder of this subsection we discuss the interpretation in detail.

The parameters $\alpha_1$ and $\alpha_2$ are defined as ratios between a flux and a concentration and thus have the dimension of an inverse time. They represent the respective time scales of the two coupled differential equations and can be interpreted as the inverse lifetime of the respective cell type. Since the average life span of osteoblasts ($\approx$3 months) exceeds the life span of osteoclasts ($\approx$2 weeks) by a factor close to 6 \cite{Manolagas00}, it is reasonable to assume that $\alpha_1 / \alpha_2 \approx 1/6$. Since the scale by which time is measured is arbitrary, we are free to fix $\alpha_1=1$.

As can be seen from Eq.~(\ref{eq:elastidef}), the remaining parameters in the Jacobian are logarithmic derivatives of the original functions. We denote these parameters as \emph{elasticities}, using a term from Metabolic Control Analysis \cite{Fell92}.
In the simple case of a linear functional dependency, the corresponding elasticity is exactly $1$. In the case of a power-law function depending on a variable $x$, $f(x)=a x^b$, the elasticity is the exponent $b$.
We note that in the model proposed in Ref.~\cite{Komarova03} all processes were modeled as power-laws. Therefore the Jacobian that is derived in this earlier paper is mathematically identical to Eq.~(\ref{eq:jacobian2d}). Nevertheless the Jacobian derived in the present work describes a larger class of models, in which processes are modeled by arbitrary positive functions.
For modeling dependencies that saturate for high concentrations, a Michaelis-Menten-type function is often assumed. In this case, the corresponding elasticity is confined to the interval $[0,1]$. It is close to $1$ in the initial region of steepest slope and approaches $0$ in the regime close to saturation.

If the exact functional form is not known, it is possible to assign a range of plausible values to the elasticities that is based on biological knowledge.
Elasticities that are associated with an activating influence (positive feedback) are positive, whereas elasticities associated with an inhibiting  influence (negative feedback) are negative. For instance, many reasonable feedback functions, such as the Michaelis-Menten function from enzyme kinetics, first grow linearly with the argument but then approach saturation for large values of the argument. For such a function, the corresponding elasticity parameter is restricted to the interval $[0,1]$ \cite{Gross04}.

In the following we assume that the osteoblasts' lifetime is not affected by additional regulators. Therefore, the decay term of osteoblasts is linear in $b$, and is independent of $c$. This translates into $g_{\rm c}=0$ and $g_{\rm b}=1$. Likewise, the decay term of osteoclasts is not influenced by osteoblasts, corresponding to $k_{\rm b}=0$. Moreover, there is no evidence for strongly nonlinear autocrine regulation of osteoblasts ($f_{\rm b}\approx 0$). Nevertheless, we analyze the bifurcations with respect to $f_{\rm b}$ in order to determine if a positive or negative feedback mechanism would affect the dynamics.

The parameters $f_{\rm c}$, $h_{\rm c}$ and $k_{\rm c}$ depend on the growth factor TGF$\beta$, an important regulator in bone remodeling \cite{Janssens05}.
When osteoclasts resorb bone tissue, TGF$\beta$ is released into the bone matrix, where it facilitates the differentiation of osteoblast progenitors to active osteoblasts, leading to $f_{\rm c}>0$.

The autocrine roles of TGF$\beta$, described by the parameters $h_{\rm c}$ and $k_{\rm c}$, are less clear:
\textit{In vitro} experiments have led to contradictory results on the influence of TGF$\beta$ especially on osteoclasts, finding both activation and repression \cite{Janssens05}. Results depend strongly on the experimental setup, such as TGF$\beta$ concentration or whether isolated cultures or co-cultures are used. According to a current view \cite{Janssens05}, TGF$\beta$ indirectly acts as a repressor by interaction with the OPG/RANKL/RANK pathway in co-cultures with osteoblasts. In isolated cultures however, TGF$\beta$ activates and sustains osteoclasts. For covering both cases, we allow the corresponding parameter, $h_{\rm c}$, to be positive or negative.

Without any feedback mechanisms, one would assume that the decay term of osteoclasts, $k_{\rm c}$, were equal to one. However, the apoptotic decay of osteoclasts has been reported to both be promoted \cite{Hughes96, Murakami98, Houde09} and suppressed \cite{Fuller00,Ruan10} by TGF$\beta$, corresponding to $k_{\rm c}>1$ and $k_{\rm c}<1$, respectively. Based on these conflicting experimental results, we assume that the corresponding elasticity $k_{\rm c}$ is positive but do not assume a specific value. 

Because of the form of the Jacobian, Eq.~(\ref{eq:jacobian2d}), only the difference 
$m_{\rm c} \equiv h_{\rm c} - k_{\rm c}$ 
is important for the stability analysis. Therefore, the effects of autocrine regulation in the production and decay terms of osteoclasts can be covered by a single parameter.

%Fig1
\begin{figure}[!ht]
\includegraphics[width=7cm]{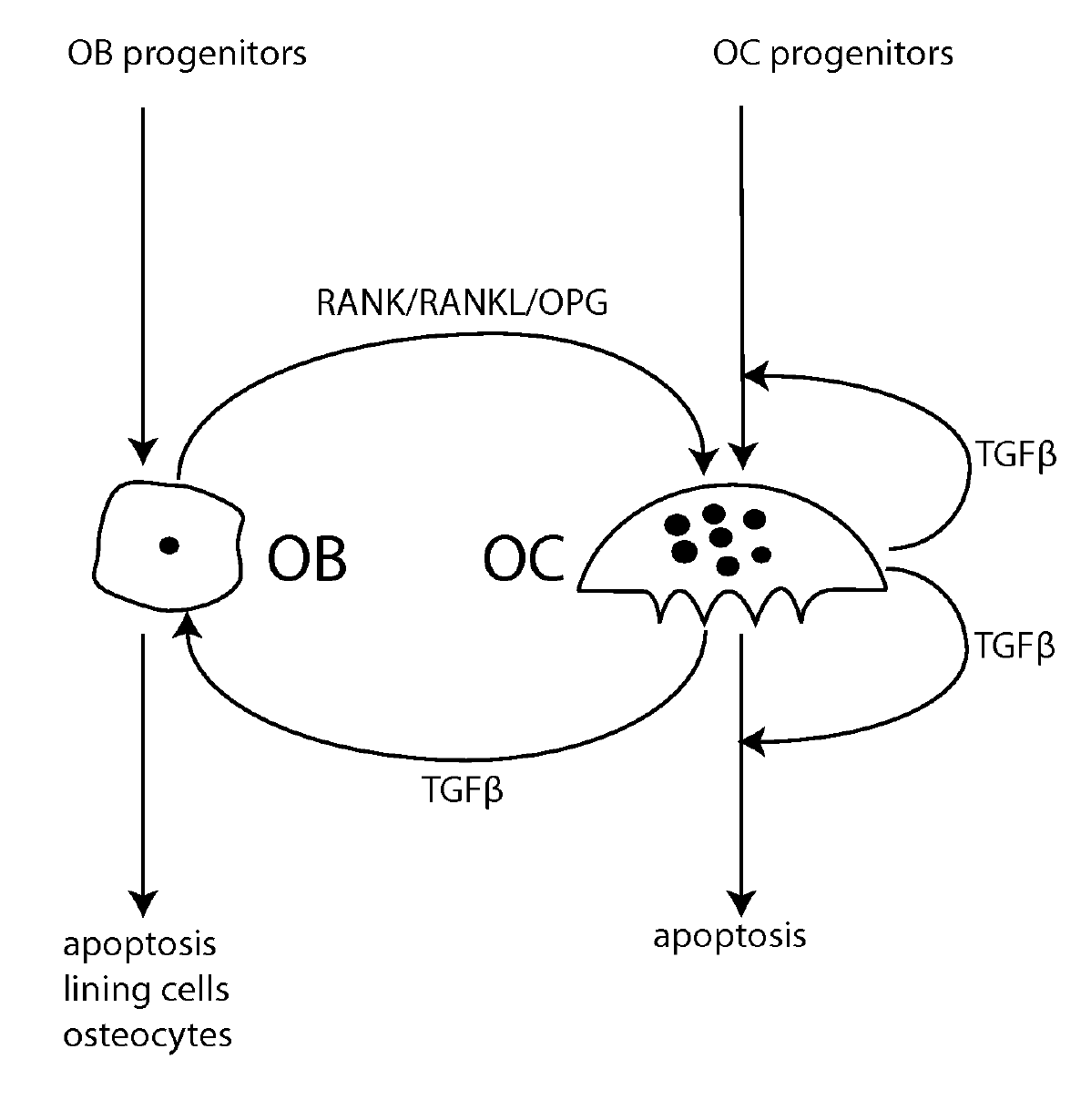}
\caption{
Schematic sketch of the two-variable model. Osteoblasts (OB) influence osteoclasts (OC) via the RANKL/RANK/OPG pathway, while the TGF-$\beta$ pathway exerts a positive feedback from osteoclasts to both osteoclasts and osteoblasts.
}
\label{g:simpscheme}
\end{figure}

Finally, the effects of the RANKL/RANK/OPG system are condensed into the parameter $h_{\rm b}$. This parameter can be negative or positive, depending on whether the repressive influence of OPG or the activation of RANK dominates.

In summary, the considerations above lead us to the Jacobian
\begin{equation}
\label{eq:jacobian2dfilled}
\rm \bf J=\begin{pmatrix} 1 & 0 \\ 0 & 6  \end{pmatrix}
\begin{pmatrix} 
f_{\rm b}-1 & f_{\rm c} \\ 
h_{\rm b} & m_{\rm c} \end{pmatrix}.
\end{equation}
with three remaining free parameters.

\subsection{Three-variable model}
\label{sec:3varintro}

It can be argued that the two variable model proposed above is oversimplified because the dynamics of precursor populations is neglected. In particular, one might miss important information by describing the population of osteoblasts with a single dynamic variable when osteoblasts have different properties at different stages of maturation \cite{Gori00, Thomas01}.
A model with a different structure has been proposed in \cite{Lemaire04} and was subsequently extended in \cite{Pivonka08, Pivonka10}. In this model, cells of osteoblastic lineage are represented by two dynamic variables, responding osteoblasts (ROBs), $R$ and active osteoblasts (AOBs), $B$. ROBs are committed to the osteoblastic lineage and interact with osteoclasts but are not yet functional osteoblasts. There are two important reasons for distinguishing AOBs and ROBs: First, there is experimental evidence that osteoblastic cells express RANKL and OPG differently at different stages of maturation, where at later stages the ratio of RANKL to OPG seems to decrease \cite{Gori00,Thomas01}. Second, TGF$\beta$, which is released and activated by osteoclasts, activates osteoblast differentiation only at early stages of differentiation, whereas it seems to enlarge the pool of responding osteoblasts by inhibiting further differentiation into active osteoblasts \cite{Janssens05}. 

%Fig2
\begin{figure}[!ht]
\includegraphics[width=7cm]{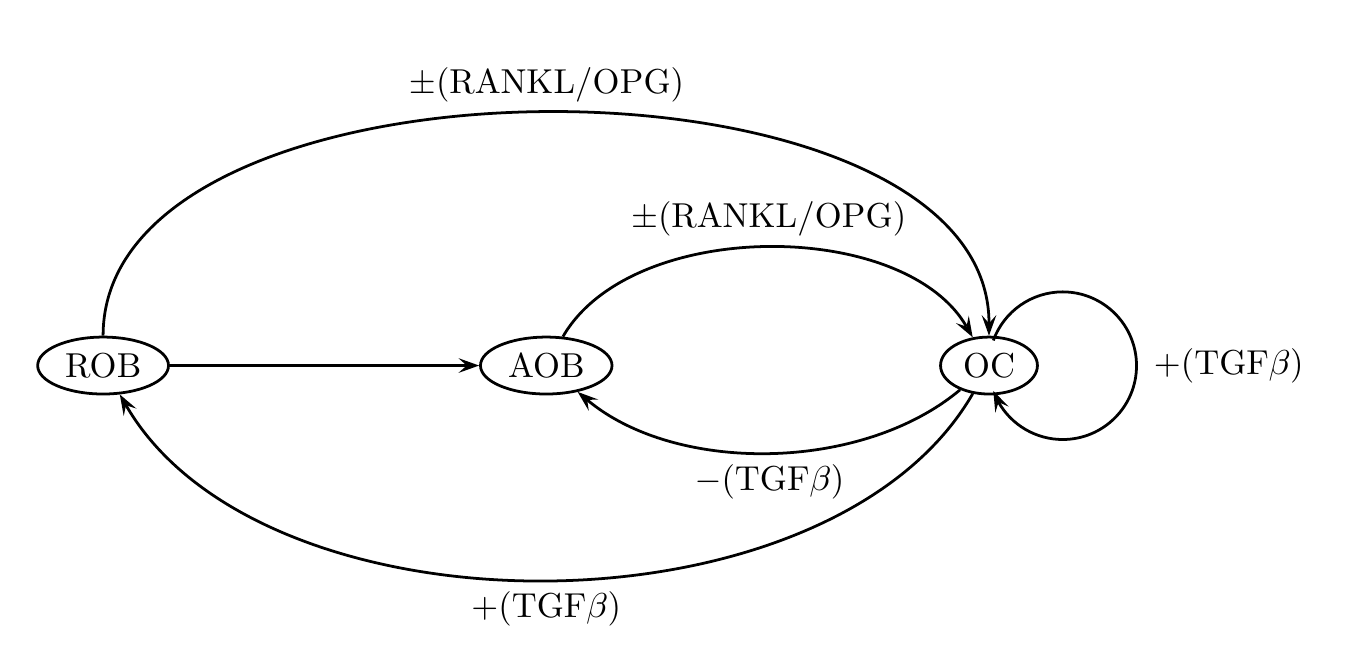}
\caption{Schematic overview of the three-variable model, in which the dynamics of responding osteoblasts (ROB), active osteoblasts (AOB) and osteoclasts (OC) is described. The feedback mechanisms, mediated by the RANK/RANKL/OPG-pathway and by TGF$\beta$, are inscribed in the diagram in the form of arcs with arrows. The straight arrow from ROB to AOB indicates a flow of biomass due to differentiation of ROBs.}
\label{g:medscheme}
\end{figure}

The structure of the three-dimensional model, shown in Fig.~\ref{g:medscheme} translates to the equations 
\begin{eqnarray*}
\frac{\mathrm{d}}{\mathrm{d} \mathrm{t}}R &=& S(C) - T(R,C) \\
\frac{\mathrm{d}}{\mathrm{d} \mathrm{t}}B &=& T(R,C) - U(B) \\
\frac{\mathrm{d}}{\mathrm{d} \mathrm{t}}C &=& V(B,R) - W(C),
\end{eqnarray*}
where the two terms in each line again correspond again to gains and losses of the population of the respective cell type.
The functional dependence in these equations is motivated by the biological processes that are included in the model. We explain these processes in more detail after formally constructing the Jacobian.

Performing the normalization procedure that was outlined in the previous section and in \cite{Gross06}, we can describe the structure of the model with the normalized equations
\begin{eqnarray*}
\frac{\mathrm{d}}{\mathrm{d} \mathrm{t}}r &=& \alpha_1 \left( s(c) - t(r,c) \right) \\
\frac{\mathrm{d}}{\mathrm{d} \mathrm{t}}b&=& \alpha_2 \left( t(r,c) - u(b) \right) \\
\frac{\mathrm{d}}{\mathrm{d} \mathrm{t}}c &=& \alpha_3 \left(v(b,r) - w(c)\right)
\end{eqnarray*}
where, in analogy to our treatment of the two-variable model, the lower-case variables and functions denote the normalized quantities and $\alpha_1$, $\alpha_2$, $\alpha_3$ are the characteristic timescales of ROB, AOB, and osteoclast turnover.

In analogy to Eq.~(\ref{eq:jacobian2d}), the Jacobian for the three-variable model can now be written as
\begin{equation}
\rm \bf J=\begin{pmatrix} \alpha_1 & 0 & 0\\ 0 & \alpha_2 &0\\ 0 & 0 & \alpha_3 \end{pmatrix}
\begin{pmatrix} 
-t_{\rm r} & 0 & s_{\rm c} - t_{\rm c} \\
t_{\rm r} & -u_{\rm b} & t_{\rm c} \\ 
v_{\rm r} & v_{\rm b} & -w_{\rm c}
\end{pmatrix}.
\label{eq:jacobian3d}
\end{equation}
A summary of the elasticities occurring in the Jacobian and the ranges we assign to them is given in Table 1.%\ref{tab:3expar}. 

The elasticities $s_{\rm c}$, $t_{\rm c}$ and $w_{\rm c}$ describe the nonlinearities in the TGF$\beta$ pathway. This pathway stabilizes the reservoir of AOBs both by promoting the differentiation of osteoblast progenitors to ROBs, leading to $s_{\rm c}>0$ and by inhibiting the further differentiation of ROBs to AOBs, leading to $t_{\rm c}<0$.
In our model, we restrict $s_{\rm c}$ to the interval $[0,1]$ and  $t_{\rm c}$ to $[-1,0]$. This range includes the choice of Hill functions with exponents equal to $1$ that were used in earlier models such as \cite{Pivonka10}.
The nature of autocrine regulation of osteoclasts has not been ultimately clarified. Therefore we restrict $w_{\rm c}$ to $[0.5,1.5]$. This range is centered around $w_{\rm c}=1$, because without any additional feedback, a linear decay term would be expected. In particular, the parameter is smaller (greater) than one if the additional feedback is negative (positive). We note that the functional forms that were assumed in Ref. \cite{Lemaire04, Pivonka08} lead to superlinear decay ($w_{\rm c} >1$).

The regulation of osteoclasts by cells of osteoblastic lineage is mediated by the RANKL/RANK/OPG pathway. 
Depending on the ratio between RANKL and its decoy receptor OPG, the corresponding elasticities $v_{\rm r}$ and $v_{\rm b}$ can be either positive or negative.
This includes all possible combinations of RANKL and OPG expression at responding osteoblasts or active osteoblasts. 
In particular, two fundamentally different scenarios are described in the literature, which are for instance discussed as models M1 and M2 in Ref.~\cite{Pivonka08}.
These scenarios are characterized in the general model by
\begin{enumerate}
 \item $v_{\rm r}<0,\quad v_{\rm b}>0$. OPG is expressed by responding osteoblasts, RANKL is expressed by active osteoblasts.
 \item $v_{\rm r}>0,\quad v_{\rm b}<0$. RANKL is expressed by responding osteoblasts, OPG is expressed by active osteoblasts.
\end{enumerate}
Intermediate situations with a differential expression of OPG and RANKL without the assumption of exclusive expression are also covered by our description (e.g. $v_{\rm r}>v_{\rm b}>0$).

\section{Results}
In this section, we show results for the two models that were introduced in the preceding section. 
The bifurcations in both models can be computed analytically from the Jacobians. In the two-variable model, the results of the bifurcation analysis depending on three parameters can be directly visualized and understood intuitively. However, in the three-variable model the larger number of parameters complicates gaining intuitive understanding. For our initial exploration of the generalized model we therefore resort to a numerical procedure which provides results that are more easily interpretable. 

\subsection{Bifurcation analysis of the two-variable model}
\label{sec:2varres}

In any system, a necessary condition for a saddle-node bifurcation is
$\det \rm \textbf{J}=0$, guaranteeing a zero eigenvalue. For the Jacobian derived in Eq.~(\ref{eq:jacobian2d}) it follows that
\begin{equation}
m_{\rm c}(f_{\rm b}-1)-f_{\rm c}h_{\rm b}=0
\label{eq:FoldCond}
\end{equation} 
has to be satisfied at a saddle-node bifurcation ($m_{\rm c}=h_{\rm c}-k_{\rm c}$).

In the Hopf bifurcation, there are two conjugate eigenvalues with zero real part. In a two-dimensional system, this means that the eigenvalues add up to zero and the trace of the Jacobian vanishes ($\mathrm{Tr} {\bf J}=0$) so that
\begin{equation}
\frac{\alpha_1}{\alpha_2}(f_{\rm b}-1) + m_{\rm c}=0.
\label{eq:HopfCond}
\end{equation}
is a necessary condition for the Hopf bifurcation.
Additionally, the inequality $\det \textbf{J} >0$ must be fulfilled.

%Fig3
\begin{figure}[!ht]
\includegraphics[width=7cm]{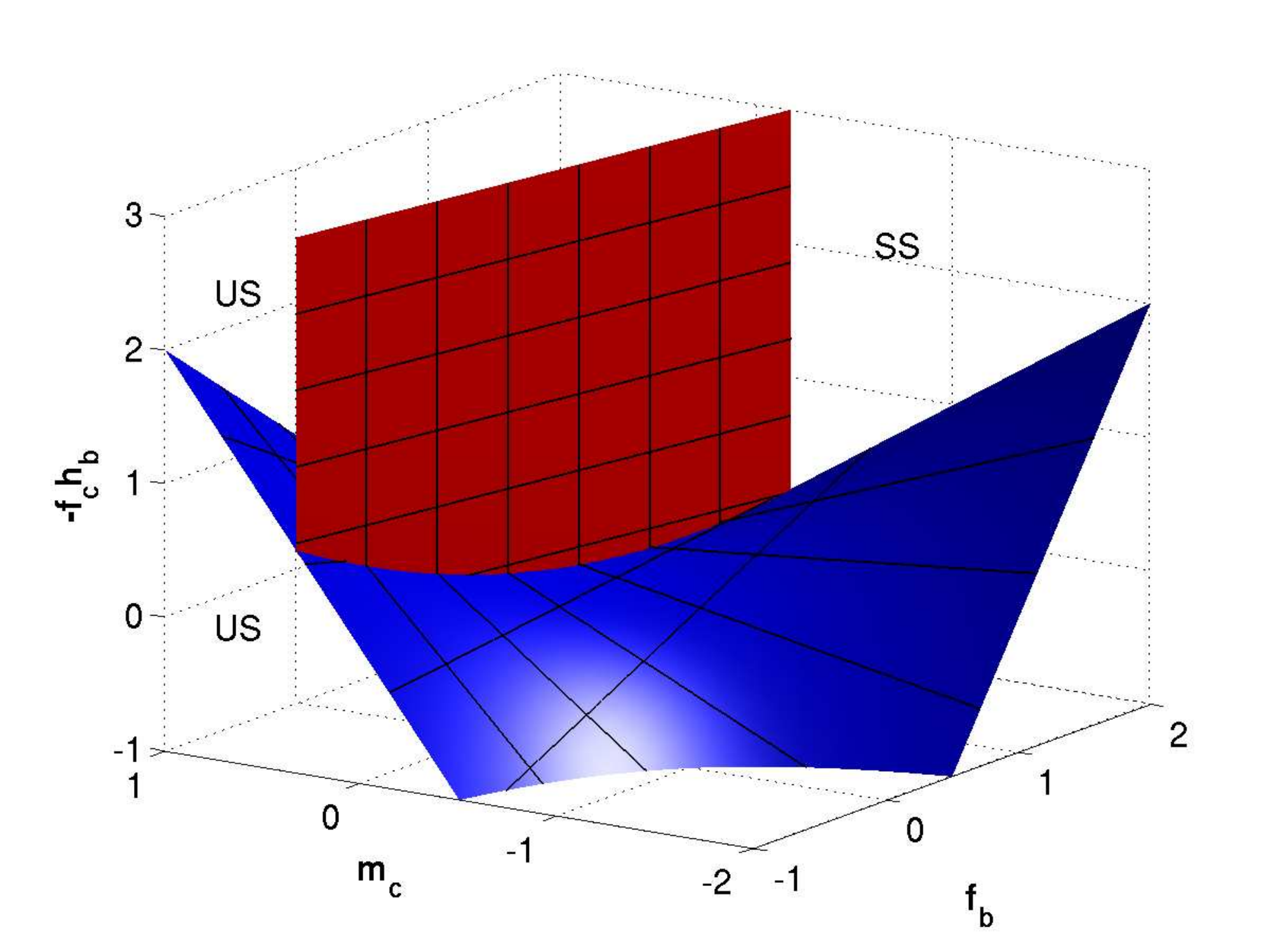}
\caption{Bifurcation diagram for the 2-variable model, depending on $f_{\rm b}$, $m_{\rm c}$ and  $-f_{\rm c}h_{\rm b}$. Each combination of the parameters in the three-dimensional volume corresponds to the steady state in a particular model. There are two distinct bifurcation surfaces that divide the regions where steady states are stable (SS) from regions where steady state unstable (US). The red surface is formed by Hopf bifurcation points, whereas the blue surface is formed by saddle-node bifurcation points.}
\label{g:bifdia2}
\end{figure}

Because the ratio of timescales $\alpha_1/\alpha_2$ is known, the bifurcation conditions, Eq.~(\ref{eq:FoldCond}) and Eq.~(\ref{eq:HopfCond}), depend only on different combinations of the three parameters, $f_{\rm b}$, $m_{\rm c}$ and  $-f_{\rm c}h_{\rm b}$ (for stability, only the product appearing in the Jacobian is important). The stability of all steady states in the whole class of models can therefore be visualized in the single three-parameter bifurcation diagram, displayed in Fig.~\ref{g:bifdia2}.

The figure shows that parameter regimes of instability can be reached both via a Hopf or a saddle-node bifurcation.
Because $f_{\rm c}>0$, we see that a high value of $h_{\rm b}$, corresponding to the case where activation by RANKL dominates over repression by OPG, destabilizes the steady state in a saddle-node bifurcation (lower region of Fig.~\ref{g:bifdia2}), while a negative $h_{\rm b}$ has a stabilizing effect. In the model, stability is therefore promoted when OPG dominates over RANKL and the effective action of the RANKL pathway is inhibiting. It is not clear whether this requirement is fulfilled \textit{in vivo}. 

The parameter regime close to the Hopf bifurcation can only be reached when $m_{\rm c} \approx 0$, i.e., when the autocrine feedback of the osteoclasts acts in an activating way as a countereffect to the linear contribution of the decay term.

 We further note that the bifurcation diagram differs from Fig.~4 a) in Ref.~\cite{Komarova03}, in which the saddle-node bifurcation surface seems to be independent from $f_{\rm b}$ (called $g_{22}$ there), which is incompatible with the form of Eq.~\ref{eq:FoldCond}.

In the generalized model we cannot determine wether the Hopf bifurcation is subcritical or supercritical without making further assumptions. We therefore consider the model proposed in \cite{Komarova03} as a specific example of the more general class considered here.
In our notation this model can be written as
 \begin{eqnarray}
 &\frac{\mathrm{d}}{\mathrm{d} \mathrm{t}}{} B = A_b B^{f_{\rm b}} C^{f_{\rm c}} - D_b B \\
 &\frac{\mathrm{d}}{\mathrm{d} \mathrm{t}} C = A_c B^{h_{\rm b}} C^{h_{\rm c}} - D_c C,  
 \end{eqnarray}
where $A_b$,$D_b$,$A_c$ and $D_c$ are rate constants
The Hopf bifurcation condition in this model is
\begin{equation}
\frac{D_c}{D_b}(h_{\rm c}-1) + (f_{\rm b}-1)=0.
\label{eq:HopfBifSpecial}
\end{equation}
This equation is equivalent to Eq.~A5 in \cite{Komarova03}.

However, the very same condition guarantees that the flow is Hamiltonian, i.e., that a function of the dynamic variables exists which is conserved on all trajectories.
We found this function to be

\begin{equation}
\begin{split}
&H(B,C)=-\frac{D_c}{f_b - 1}B^{1-f_b} C^{1-h_c} \\ 
&- \frac{A_c}{h_{b}- f_b + 1} B^{h_{b} - f_b + 1}
 + \frac{A_b}{f_{c} - h_c + 1} C^{f_{c} - h_c + 1}
\end{split}
\end{equation}
It can easily be verified by direct differentiation that under the condition of Eq.~(\ref{eq:HopfBifSpecial}), $\frac{\mathrm{d} H}{\mathrm{d} t}=0$. The actual Hamilton equations are fulfilled after the coordinate transformation $p(B)=\frac{1}{1-f_b} B^{1-f_b}$ and $q(C)=\frac{1}{1-h_c} C^{1-h_c}$.

It follows that no limit cycle with a defined amplitude is born in the Hopf bifurcation and the steady state is a center. The Hopf bifurcation is neither subcritical nor supercritical, but is just at the brink between the two alternatives. This structural instability is caused by a symmetry in the model. Sustained oscillations occur only exactly at the bifurcation point and the amplitude of the oscillations will depend strongly on initial conditions.

The symmetry that causes the degenerate behavior of the model is broken if the autocrine regulation of the osteoclast removal is taken into account ($k_{\rm c} \neq 1$), as it has been assumed in other models \cite{Lemaire04}. Both for negative and positive autocrine feedback, the system is no longer Hamiltonian at the Hopf bifurcation, and depending on the actual parameters the bifurcation is subcritical or supercritical. We verified numerically that close to supercritical Hopf bifurcations, stable limit cycles with a well-defined amplitude can exist and sustained oscillations are possible over a wider range of parameters. These findings imply that the two-dimensional mathematical model is sensitive with respect to the existence of feedback in the removal term of osteoclasts..

In summary, the bifurcation analysis shows that in the two-dimensional model structure, it is necessary for stability that the feedback exerted by osteoblasts on osteoclasts is effectively inhibiting, which is the case when the repressing effects of OPG dominate over the activating effects of RANKL. Otherwise, the steady state under consideration is unstable and therefore cannot correspond to a physiological equilibrium. Furthermore, we showed that a Hopf bifurcation exists close to the realistic parameter regime. A pathological shift in the parameters may therefore drive the system over the Hopf bifurcation, which will typically lead to stable sustained oscillations of osteoclast and osteoblast numbers if the osteoclast removal rate increases at least weakly with the number of osteoclasts. 

\subsection{Bifurcation analysis of the three-variable model}
In the three variable model we cannot visualize all factors impinging on stability in a single diagram. We therefore use a random sampling procedure \cite{Steuer06} for gaining a first impression of the effect of the various parameters. This analysis is then combined with a bifurcation analysis of three-dimensional subspaces of the larger parameter space.

%Fig4
\begin{figure}[!ht]
\includegraphics[width=9cm]{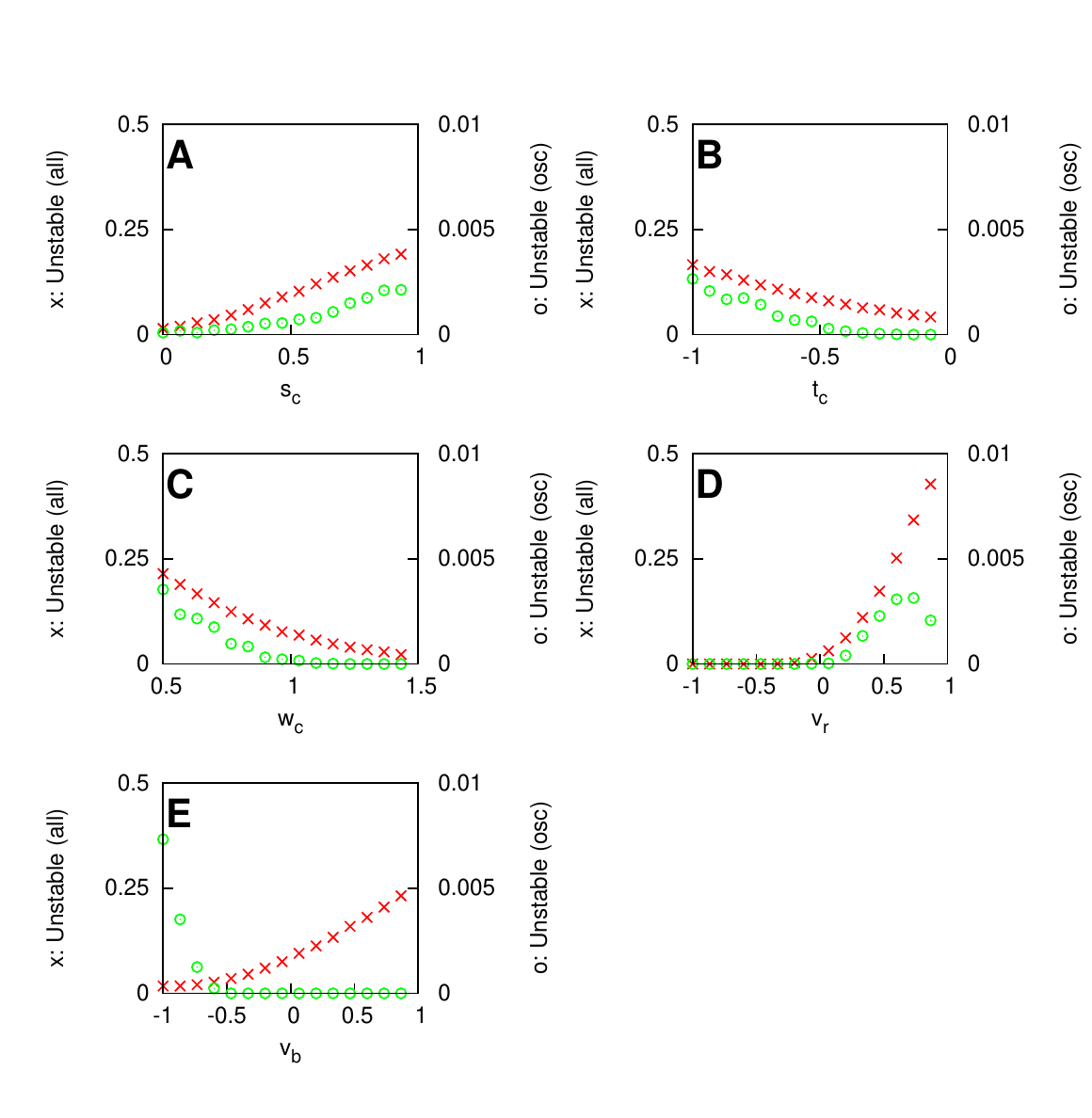}
\caption{
Effect of parameters on local dynamics. Using an ensemble of $10^6$ randomly drawn steady states, the histograms show the percentage of randomly drawn states that are unstable (red crosses) and the fraction of randomly drawn states which are unstable \emph{and} have leading complex eigenvalues, indicating oscillatory instabilities (green circles). Each panel shows the effect of one of the elasticities, while averaging out the effect of the other parameters. An ascending curve signifies that for values at the top of the range of the respective parameter, steady states are more likely to be unstable than for low values of the parameter, while a descending curve signifies the opposite.
}
\label{g:stabexpar3}
\end{figure}

For the three-variable model, we performed a statistical sampling sampling analysis by creating $N=10^7$ random parameter sets. In each set, we assigned to all parameters random values that were drawn from uniform distributions, using the intervals defined in Table \ref{tab:3expar}. 

%Tab1
\begin{table}[!ht]
\begin{tabular}{c c c c}
Parameter & Interpretation & Range\\
\hline
$s_{\rm c}$ & activation of ROB production & $[0,1]$ \\
$t_{\rm r}$ & ROB decay, linear in $r$ & $1$ \\
$t_{\rm c}$ & repression of ROB decay & $[-1,0]$ \\
$u_{\rm b}$ & AOB decay, linear in $b$ & $1$ \\
$v_{\rm r}$ & action of ROBs on OC & $[-1,1]$ \\
$v_{\rm b}$ & action of AOBs on OC & $[-1,1]$ \\
$w_{\rm c}$ & activation of OC decay & $[0.5,1.5]$ \\
\end{tabular}
\caption{Parameters in the three-variable model}
\label{tab:3expar}
\end{table}

We then determined the stability of the steady state in each sample by numerically computing the eigenvalues of the Jacobian. Based on this ensemble of random steady states, we then statistically analyse the relations between the parameters and stability by creating histograms for each parameter, showing the percentage of unstable steady states in the whole ensemble as a function of the parameter value.
The results of this analysis are shown in the histograms in Fig.~\ref{g:stabexpar3}. 

Panel A and B of Fig.~\ref{g:stabexpar3} show that a strong paracrine effect of TGF$\beta$ on osteoclasts  ($s_{\rm c}\gg 0$ and $t_{\rm c} \ll 0$) destabilizes steady states (i.e., on the left side of Fig.~\ref{g:stabexpar3}A where $s_{\rm c}$ is small, the proportion of randomly generated states that are unstable is close to zero).
The parameter $w_{\rm c}$ has the opposite effect (Fig.~\ref{g:stabexpar3}C): Strong positive feedback of osteoclasts on osteoclast removal stabilizes the steady state. It follows that the paracrine effects of TGF$\beta$ on osteoblasts that are described by  $s_{\rm c}$ and $t_{\rm c}$ destabilize the steady state, whereas the apoptosis-inducing autocrine effects of TGF$\beta$, described by $w_{\rm c}$, stabilize it.

For $v_{\rm r}<0$, we detected few unstable states (Fig.~\ref{g:stabexpar3}D), showing that models in which only OPG is preferentially expressed on ROBs usually operate from a stable steady state that cannot be destabilized easily. 
The other parameter that is related to RANKL signaling, $v_{\rm b}$, also acts destabilizingly at large positive values (Fig.~\ref{g:stabexpar3}E).
As noted above, there is experimental evidence that OPG is expressed stronger on active osteoblasts, while RANKL is expressed stronger on ROBs. This implies that the parameter regime that is most likely realized in nature is characterized by $v_{\rm r}>v_{\rm b}$, which is also the regime in which instabilities occur most frequently.

In order to investigate the nature of the instabilities in more detail, we distinguished between unstable steady states in which the eigenvalue with the largest real part is a real number and those in which it is part of a complex conjugate pair. The significance of this eigenvalue lies in its effect on the departure of the system from the unstable state. Specifically, when departing from a state in which the eigenvalue with the largest real part has a non-zero imaginary part, the system launches into oscillations.

Figure~\ref{g:stabexpar3} shows that for most parameters the fraction of unstable states with a complex leading eigenvalue changes proportionally to the total fraction of unstable states for most parameters. However, very different behavior is observed for the parameter $v_{\rm b}$ (Fig.~\ref{g:stabexpar3}E):
While the fraction of unstable states with a real positive eigenvalue increases with increasing $v_{\rm b}$, the fraction of unstable states with a leading pair of complex conjugate eigenvalues decreases with an increasing $v_{\rm b}$.
This behavior suggests that the main route to instability is via a saddle-node bifurcation, but the probability for encountering a Hopf bifurcation increases with decreasing $v_{\rm b}$.

Now that we have identified the overall impact of the parameters on stability, we proceed by investigating selected parameters in bifurcation diagrams. Here, we chose to concentrate on the parameters $v_{\rm b}$ and $v_{\rm r}$ for which the random-sampling analysis turned out interesting results, as well as the parameter $w_{\rm c}$, which captures the different ways in which osteoblast removal has been modeled in earlier models.

%Fig5
\begin{figure}[!ht]
\includegraphics[width=7cm]{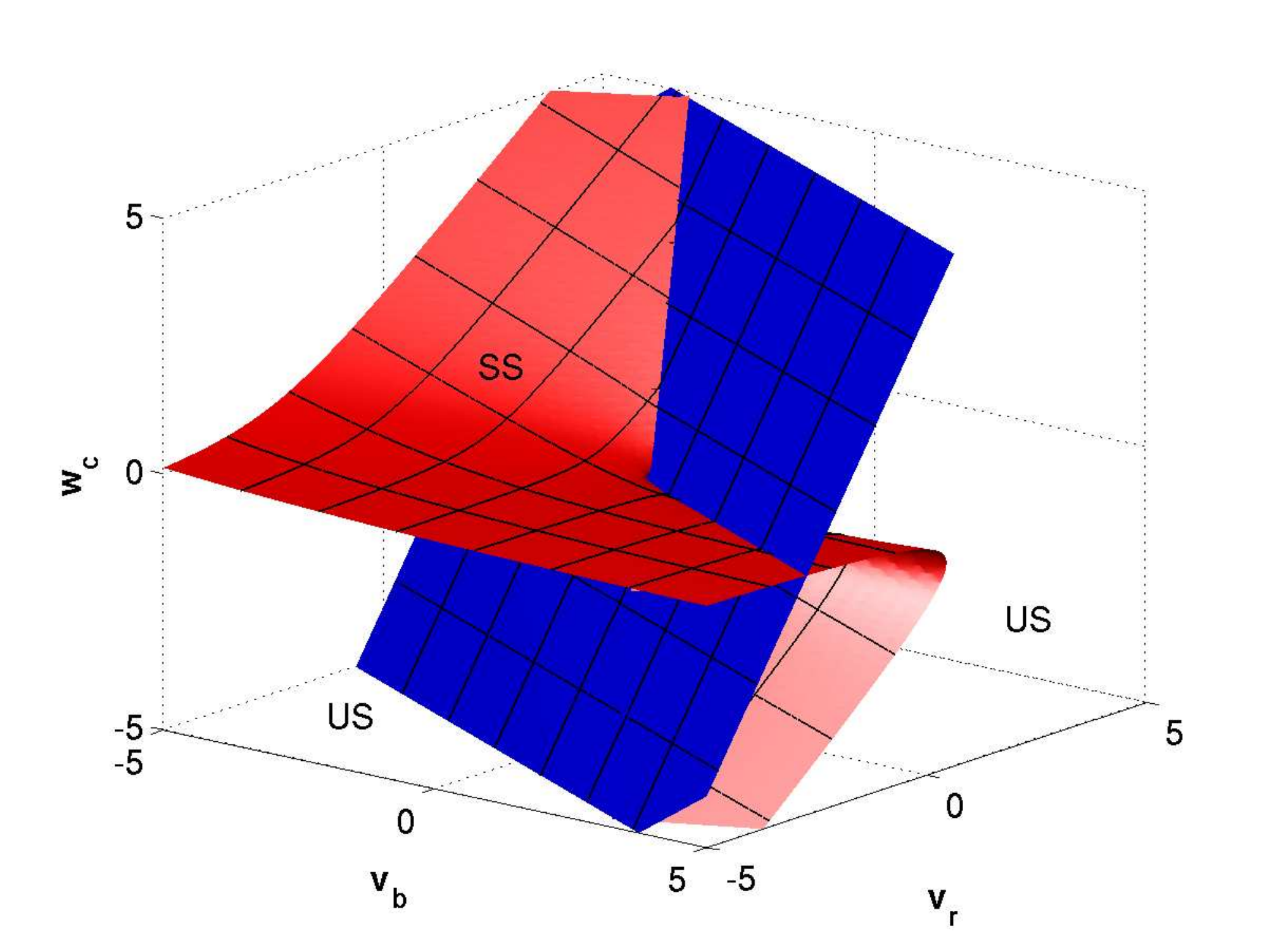}
\caption{Bifurcation diagram of the extended model, depending on the effects of RANKL/OPG ($v_{\rm b}$, $v_{\rm r}$) and the autocrine effects of osteoblast decay ($w_{\rm c}$). The stable parameter regime (SS), which is located in the upper front part of the diagram, can be lost via a Hopf bifurcation (red) or a saddle-node bifurcation (blue). Other parameters: $\alpha_1=1$, $\alpha_2=1$, $\alpha_3=6$, $s_{\rm c}=0.8$, $t_{\rm c}=-0.8$.}
\label{g:bif3dlarge}
\end{figure}

The bifurcation diagram in Fig.~\ref{g:bif3dlarge} shows that parameter sets corresponding to stable steady states are characterized by large values of $w_{\rm c}$ and small values of $v_{\rm r}$ (upper front of the figure).
The section in parameter space that is most likely realized in nature based on experimental results is characterized by $w_{\rm c} \approx 1$, which can be close to a Hopf bifurcation depending on the values of $v_{\rm r}$ and $v_{\rm b}$ that describe whether OPG and RANK are expressed preferentially on osteoblast precursors or active osteoblasts.
Moreover, Fig.~\ref{g:bif3dlarge} confirms the findings from the random-sampling analysis that both large values of $v_{\rm r}$ and small values of $w_{\rm c}$ act in a destabilizing way. 

For the parameter $v_{\rm b}$, the situation is more complicated: For large values of $v_{\rm b}$, the steady state loses its stability in a saddle-node bifurcation, whereas stability is lost in a Hopf bifurcation for small values. This explains that the stability curve for the parameter $v_{\rm b}$ in Fig.~\ref{g:stabexpar3}D depends on whether  all unstable states or only those with an oscillatory instability were taken into account. The minimum observed for $v_{\rm r}$ in Fig.~\ref{g:stabexpar3} can be explained by the saddle-node bifurcation surface replacing the Hopf bifurcation surface as the primary source of instability when $v_{\rm r}$ is increased.

We note that the bifurcation diagram in Fig.~\ref{g:bif3dlarge} contains several bifurcations of higher codimension. While the bifurcations discussed so far are of codimension $1$, forming two-dimensional planes in a three-dimensional bifurcation diagram, bifurcations of higher codimension appear as lines or points.
The Hopf-bifurcation surface ends in a Takens-Bogdanov bifurcation of codimension two as it connects to the saddle-node bifurcation surface. For low values of $w_{\rm c}$, the Hopf-bifurcation intersects the saddle-node bifurcation in a Gavrilov-Guckenheimer bifurcation. In the center of the Figure, the Takens-Bogdanov bifurcation and the Gavrilov-Guckenheimer bifurcation intersect in a triple point bifurcation.
The presence of codimension-2 bifurcations can be of relevance for applications because they can imply the existence of non-local properties such as homoclinic bifurcations or chaos. However, a detailed discussion of these implications is beyond the scope of the present paper. Instead, we refer the reader to \cite{Gross05, Kuznetsov95}.

%Fig6
\begin{figure}[!ht]
\includegraphics[width=7cm]{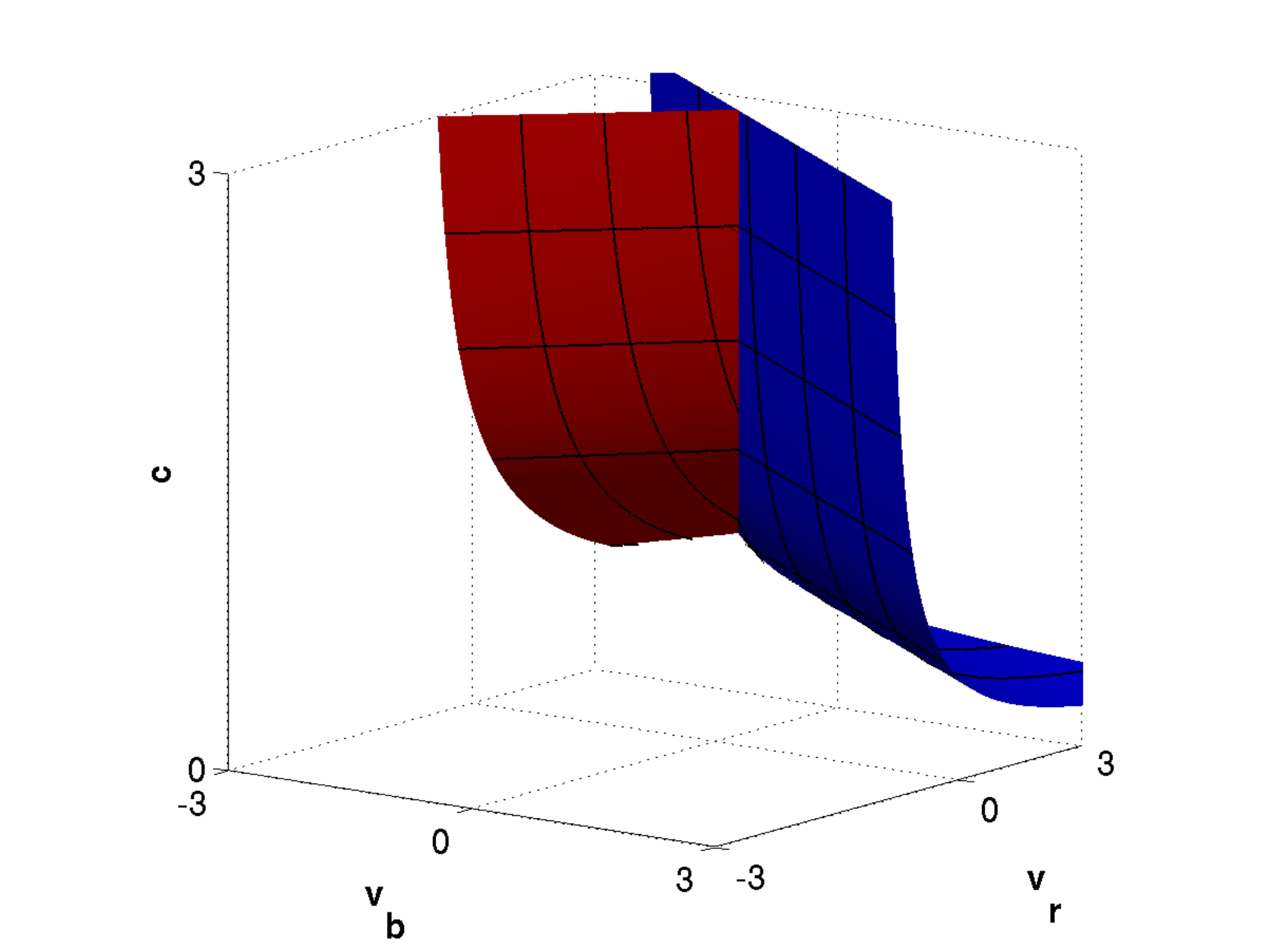}
\caption{Bifurcation diagram in which all processes controlled by TGF$\beta$ exhibit the same degree of nonlinearity. The strength of this nonlinearity is described by the parameter  $c =s_{\rm c}=-t_{\rm c}=w_{\rm c}-1$. The other bifurcation parameters, $v_{\rm r}$ and $v_{\rm b}$, describe the effect of the RANKL pathway. In the diagram, the red surface describes a Hopf bifurcation, whereas the blue surface describes a saddle-node bifurcation. In the front region of the diagram, steady states are stable. Other parameters: $\alpha_1=1$, $\alpha_2=1$, $\alpha_3=6$}
\label{g:bif3dlemaire}
\end{figure}

A different section of the parameter space was considered in Ref.~\cite{Lemaire04}, where a single Hill function with one $K_m$ value was chosen for all processes controlled by TGF$\beta$. In the case of repression, the inverse of this function was used. Translated into the framework of generalized modeling, this means that $s_{\rm c}=-t_{\rm c}=w_{\rm c}-1 = c$ and thus a reduction of free parameters. The new parameter $c$ changes simultaneously the local nonlinearity of all functions describing the TGF$\beta$ pathway. The diagram in Fig.~\ref{g:bif3dlemaire} shows that an increase of the effective feedback by TGF$\beta$ can lead to instability, which was not observed in Fig.~\ref{g:bif3dlarge} where only the parameter $w_{\rm c}$ was varied. The bifurcation properties with respect to the other parameters, $v_{\rm r}$ and $v_{\rm b}$, are consistent with Fig.~\ref{g:bif3dlarge}.

\section{Discussion}
In this paper, we have investigated the dynamics of a large class of models for bone remodeling using the approach of generalized modeling. Investigating the bifurcation behavior of a two-variable model topology, we showed that both saddle-node bifurcations and Hopf bifurcations can occur. 

In the two-dimensional model, stability of the steady state requires that the effect of OPG dominates over that of RANKL to make the two-dimensional system an adequate model for the process of bone remodeling.
We further showed the possibility of negative or positive autocrine feedback on osteoclasts should be taken into account in models because assuming a linear removal rate can lead to structurally unstable models. In such models any small deviation from the model assumptions can lead to qualitatively different behavior. Because the generalized model proposed here does not need to assume any specific functional form, it avoids such degeneracies that often are caused by an unfortunate choice of functional forms in conventional models.

In the analysis of an alternative model with three variables, we combined a random sampling approach with a bifurcation analysis of specific subclasses of models.
The three-dimensional model incorporates experimental findings suggesting that RANKL is expressed preferentially on responding osteoblasts, while its antagonist OPG is mainly expressed on matured active osteoblasts. These conditions place the system into an area of the bifurcation diagram that is close to both saddle-node and Hopf bifurcations.
The stability analysis therefore shows that in the dynamical system of bone remodeling, various bifurcations not only exist but are located in a parameter space supported by experimental findings. 

The main benefit of operating close to a region of instability is probably that a stronger adaptive response to external changes of the model parameters is possible. Although our modelling approach is not designed to study the response of the model to perturbations directly, it has been shown before by direct simulations that conventional models that fall into the general class considered here respond strongly to perturbations, thus allowing a better functional control in bone remodeling \cite{Pivonka08}.

Despite the benefits, operating close to a bifurcation also poses risks to the system.
A change in the parameters by an external process can shift the system over the bifurcation, so that the stable steady state becomes unstable or ceases to exist.
It is therefore reasonable to ask wether certain diseases of bone can be related to bifurcations, leading to qualitatively different dynamical behavior.

It is known that several diseases of bone are related to dysfunctions in the regulation of bone remodeling, among them postmenopausal osteoporosis, Paget's disease, osteopetrosis and osteopenia.
There is some evidence that diseases may lead to qualitatively different dynamical behavior. Periodic activity of osteoclasts has been observed in Paget's disease of bone \cite{Reddy01} and also \textit{in vitro} \cite{Akchurin08}. It is conceivable that these dynamics are evoked by the transition of a steady state to instability in a Hopf bifurcation. In the two-dimensional model, Hopf bifurcations are most likely caused by increasing the activating autocrine feedback of osteoclasts. Yet, TGF$\beta$ was found to be not related to Paget's disease of bone \cite{Ralston94}. In the three-dimensional model, however, stability can also be lost in Hopf bifurcations by increasing the RANKL/OPG ratio, which is in agreement with findings that the RANKL pathway is involved in Paget's disease. In particular, OPG deficiency was reported to be related to juvenile Paget's disease \cite{Whyte06}.

In conclusion, it is still unclear if known diseases of bone can be connected to bifurcation phenomena. However, the analysis presented here suggests that Hopf and saddle-node bifurcations exist close to the physiological steady state. We do not claim that specific diseases can be related to these bifurcations. Yet, a bifurcation occurring \textit{in vivo} should certainly lead to a pathological condition. Therefore it seems very plausible that a connection for instance between the crossing of a Hopf bifurcation and the onset of Paget's disease may exist. If this is indeed confirmed it would imply that powerful tools of bifurcation theory and related data analysis techniques, can be applied to explore the dynamics of the disease.

\addcontentsline{toc}{chapter}{\numberline{}Bibliography}
\bibliographystyle{elsart-num}
%\bibliography{lit_bone}

\end{document}